\documentclass[a4paper,twocolumn]{revtex4-1} 
\usepackage{amsmath}
\usepackage{graphicx}
\usepackage{hyperref}
\usepackage[utf8x]{inputenc}

\usepackage{fancyhdr}
\fancypagestyle{firstpage}{%
  \lhead{This is the pre-peer reviewed version of the following article:   \textit{Phys. Status Solidi C} \textbf{10}  (2013) 468-471, which has been published in final form at \href{https://doi.org/10.1002/pssc.201200701}{https://doi.org/10.1002/pssc.201200701}. This article may be used for non-commercial purposes in accordance with Wiley Terms and Conditions for Use of Self-Archived Versions.}

  \rhead{}
}

\begin{document}

\title{Two modes of HVPE growth of GaN and related macrodefects}


\author{  V. V. Voronenkov$^{1,2}$ } \email{voronenkov@mail.ioffe.ru}
\author{   N. I. Bochkareva\textsuperscript{2}}
\author{   R. I. Gorbunov\textsuperscript{2}}
\author{   P. E. Latyshev\textsuperscript{3}}
\author{   Y. S. Lelikov\textsuperscript{2}}
\author{   Y.~T.~Rebane\textsuperscript{2}}
\author{   A.~I.~Tsyuk\textsuperscript{2}}
\author{   A.~S.~Zubrilov\textsuperscript{2}}
\author{   U. W. Popp\textsuperscript{4}}
\author{   M. Strafela\textsuperscript{4}}
\author{   H. P. Strunk\textsuperscript{4}}
\author{   Y. G. Shreter\textsuperscript{2}} \email{y.shreter@mail.ioffe.ru}

\affiliation{  \textsuperscript{1}\,St. Petersburg State Polytechnical University, Polytechnicheskaya 29, St. Petersburg 194251, Russia\\
  \textsuperscript{2}\,Ioffe Physical Technical Institute, Polytechnicheskaya 26, St. Petersburg 194021, Russia\\
  \textsuperscript{3}\,Fock Institute of Physics, St. Petersburg State University, Ulyanovskaya 1, Petrodvorets, St. Petersburg 198904, Russia\\
 \textsuperscript{4}\,Institute of Materials Science, Chair of Material Physics, University of Stuttgart, Heisenbergstr. 3, 70569 Stuttgart, Germany}




\begin{abstract}
GaN films with thickness up to 3 mm were grown by halide vapour phase epitaxy method.
Two growth modes were observed: the high temperature (HT) mode and the low temperature (LT) mode.
Films grown in HT mode had smooth surface, however the growth stress was high and caused cracking.
Films grown in LT mode had  rough surface with high density of V-defects (pits), however such films were crack-free.
The influence of growth parameters on the pit shape and evolution was investigated.
Origins of pits formation and process of pit overgrowth are discussed.
Crack-free films with smooth surface and reduced  density of pits were grown  using combination of the LT and HT growth modes.
\end{abstract}

\maketitle   
    \thispagestyle{firstpage}

\section{Introduction}
Significant problem in GaN industry is a lack of widely available native substrates.
One of the methods of producing  thick GaN films is halide vapour phase epitaxy (HVPE).
Macrodefects, namely cracks and pits, are main problem in producing thick GaN films by HVPE .

Cracks are formed in GaN films during growth  \cite{etzkorn2001cracking} and cool down to the room temperature.
The origin of the crack formation during growth is the tensile stress, originating in growth process.
In our opinion \cite{tsyuk2011effect} the possible origin of this stress is the climb of threading dislocations~\cite{romanov2003stress}.
The transport of point defects, responsible for the dislocation climb, may occur in bulk \cite{moram2010dislocation,kuwano2011evidence} or at the surface \cite{cantu2005role,follstaedt2009strain}.

Pits are a common type of defects in epitaxially grown crystals, well known since early studies~\cite{givargizov1967two,sheftal1966magometov}.
The pits form on the surface of the film in growth process because of various imperfections such as threading dislocations~\cite{chen1998pit} or
inversion domains~\cite{lucznik2009bulk}.

In this paper we studied the growth of thick (up to 3~mm) GaN films.
We had found that at low growth temperature  the main defects were pits,
while at high growth temperature the main defects were cracks.
\section{Experimental}
GaN films with thickness from 10~$\mu$m to  400~$\mu$m were grown in a custom  6x2'' wafer vertical HVPE reactor.
GaN boules with thickness from 400~$\mu$m to 3~mm were grown in custom 1x2'' vertical reactor optimized for long growth processes.

The $c$-plane sapphire substrates were used.
The substrate surface was nitridated in ammonia atmosphere at 1060~$^{\circ}$C.
Then low-temperature GaN buffer  layer was deposited at 600~$^{\circ}$C and total pressure of 250~Torr.
The thickness of buffer layer was about 1 $\mu$m.
After that the main growth process was conducted.
The total reactor pressure during growth was 800~Torr.
Nitrogen was used as a carrier gas.

The growth rate ($V_{gr}$) was measured \textit{ex situ} by weighting the grown film and was found to be proportional to the partial pressure of  GaCl (P$_{\rm GaCl}$).

A series of more than 50 films with thickness in the range 30 -- 100 $\mu$m were grown in order to study the influence of
growth parameters on the surface morphology.
The growth temperatures ranged from  990~$^{\circ}$C to  1140~$^{\circ}$C.
The growth rate  was controlled by the flow rate of GaCl and varied from 10 $\mu$m/hour to 200 $\mu$m/hour.

The surface morphology and the internal structure of the films were examined with optical microscope.
To observe the pits optically polished cross-sections of the films were prepared.
The cut was made as near as possible to the investigated pits.
After that samples were lapped and polished to make the surface of the cut optically smooth.

\section{Results and discussion}
\begin{figure*}[]%
  \includegraphics*[width=1.0\linewidth]{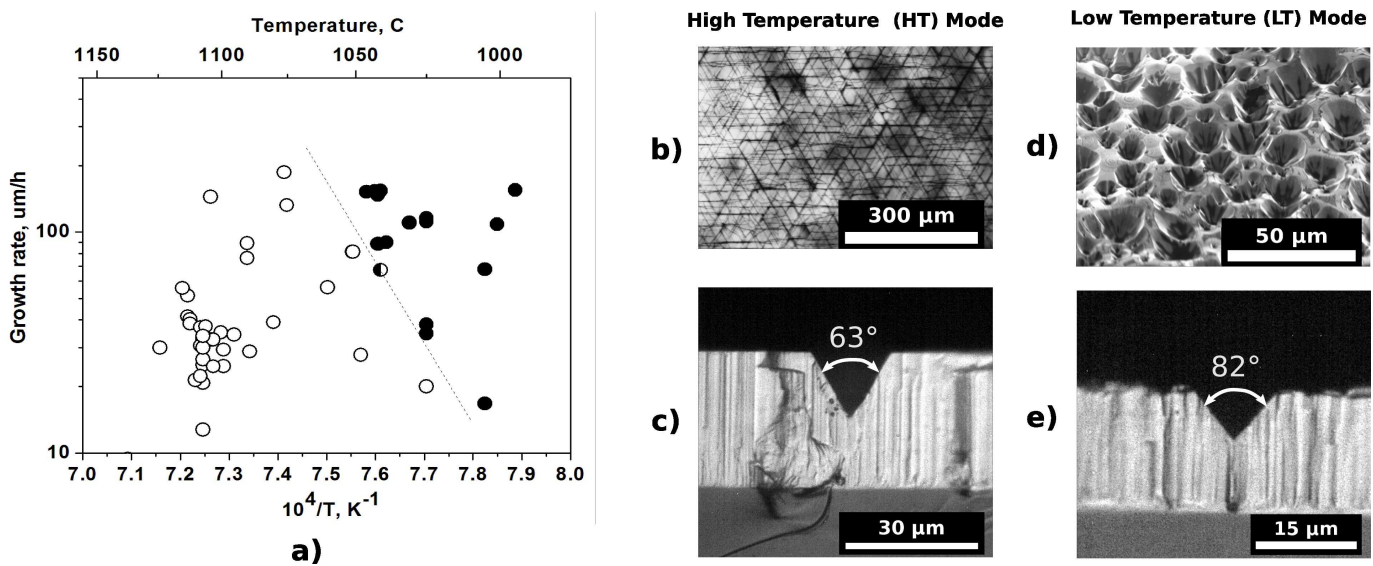}

\caption[]{Comparison of LT and HT growth modes.
a) Dependence of growth mode on growth parameters. Each point on the plot represents a film grown in a individual growth process.
Open circles correspond to a film exhibiting high temperature morphology, filled circle --  low temperature morphology,
 the black-white circle corresponds to a film shown in Fig.~\ref{LT-HT}.
The dashed line is an eye guide and represents the transition between the growth modes. b) Transmission light micrograph of film grown in HT mode.
c) Optical micrograph of the cleavage of HT film across a pit. d) Focused ion beam micrograph of film grown in LT mode. e) Optical micrograph of the cleavage of LT film across a pit.}
\label{modes}
\end{figure*}

The type of observed morphology depended on the growth parameters, mainly temperature and partial pressure of $\rm GaCl$.

Two distinct types of film morphology corresponding to different growth modes were observed.
These modes will be further referred to as a  high temperature (HT) mode and  low temperature (LT) mode. 

The dependence of the growth mode on temperature and growth rate is presented in Fig.~\ref{modes}a.
Each point on a plot represents a film grown in an individual process, at given temperature and growth rate.
Open and filled circles correspond to films having HT and LT morphology respectively.
The LT mode was observed in range T = 990 -- 1050~$^\circ$C, $V_{gr}$ = 10 -- 200~$\mu$m/hour.
The HT mode was observed in range T = 1020 -- 1140~$^\circ$C, $V_{gr}$ = 10 -- 200~$\mu$m/hour.
The regions of LT and HT modes do not overlap.
The dashed line is  an eye-guide and represents the dividing line between LH and HT mode.

Films grown in HT mode had smooth surface.
A network of buried cracks could be observed inside such films, Fig.~\ref{modes}b.
The tensile growth stress could be roughly estimated by the average crack spacing \cite{etzkorn2001cracking}.
The typical crack spacing ranged form 10 $\mu$m to 50 $\mu$m, depending on the growth parameters.
From this values the growth stress was estimated to be from 0.2 GPa to 0.4 GPa.
The density of the pits was about  1  cm$^{-2}$.
 Pits were typically hexagonal and composed of $\langle11\overline{2}2\rangle$ facets. 
 The opening angle of pits was about 63 degrees,  Fig.~\ref{modes}c.

Films grown in LT mode had rough surfaces covered by pits  and hillocks, Fig.~\ref{modes}d.
The opening angle of the pits was in range from 80 to 100 degrees, Fig.~\ref{modes}e.
The pits were cone-shaped.
Pits were distributed evenly and occupied a large part of the film surface.
The films are typically crack free that indicated a relatively low value of growth stress.

\begin{figure}[]%
  \includegraphics*[width=1.0\linewidth]{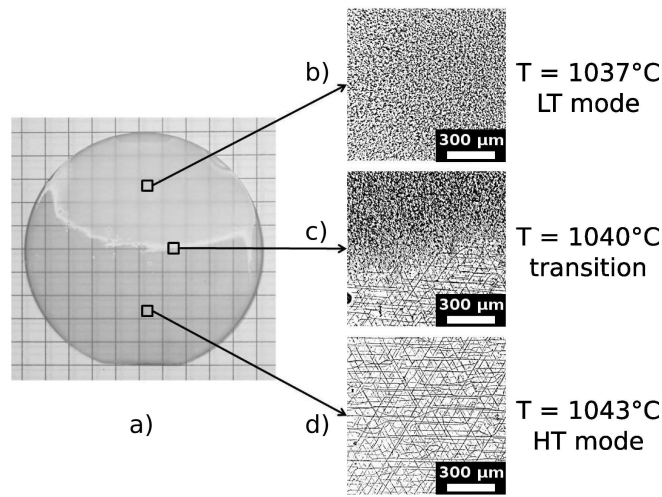}
 \caption[]{ 
   GaN film on a 2'' sapphire substrate, grown on a substrate holder with nonuniform heating.
a) Photograph of the film. The upper part of the film exhibits LT growth mode, the lower part -- HT mode.
b-d) Transmitted light micrographs of various parts of the film.
}
 \label{LT-HT}
 \end{figure}

To study the transition between the growth modes a special substrate holder with nonuniform heating was used.
The temperature across the substrate varied from 1045~$^{\circ}$C near the basal cut of the substrate, to 1035~$^{\circ}$C on the opposite side.
The film grown on this holder is shown in Fig.~\ref{LT-HT}.
The cold part of a film exhibited LT mode morphology with rough surface, high pit density and low growth stress.
The hot part had an HT mode morphology with smooth surface and high growth stress.
The transition occurred in a narrow region with width of several hundred micrometers. 
Both surface morphologies and value of growth stress change simultaneously.

\begin{figure}[]%
  \includegraphics*[width=1.0\linewidth]{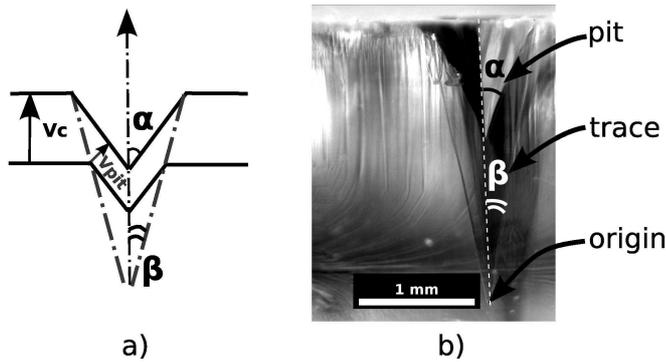}
 \caption[]{ Analysis of the evolution of pit geometry.
    a) Crystal surface with pit at two consecutive moments. Size of the pit changes according to relation of $V_{pit}$ and $V_c$.
    b) Side view of a pit grown in HT mode. A pit and it's trace is shown by arrows.
}
 \label{math}
 \end{figure}

Examination of the cross-sections of thick films showed that each pit leaves a visible trace during growth, Fig.~\ref{math}b.
These pit-related traces can be also observed by photoluminescence or cathodoluminescence studies \cite{wagner2002influence} or revealed by selective photo-etching \cite{weyher2010revealing}.
The traces permit to observe the entire history of pits, from generation to overgrowth.

The shape of the trace is determined by the relation of growth rate of the pit facet and the growth rate of the $c$-plane.
If the angles of  facets ($\alpha$) and the trace of the pit ($\beta$) are measured, the relative growth rate can be calculated:
\begin{equation}
 \frac{V_{pit}}{V_{c}} = (\tan \alpha - \tan \beta)\cos \alpha.
\label{eq-rates}
\end{equation}
It can be seen from Fig.~\ref{math}a that pits are widening with growth time when $\beta > 0$ and narrowing when $\beta <0$.
 Substitution $\beta = 0$ in Eq. \ref{eq-rates} gives condition of pit overgrowth:

\begin{equation}
 V_{pit} > V_{c}\sin\alpha.
\label{eq-criterion}
\end{equation}

According to our results and literature data \cite{sheftal1966magometov,chen1998pit,lucznik2009bulk,motoki2007dislocation} the origin of pit formation could be imperfections on initially smooth surface that locally reduce growth rate.
The defects that locally reduce growth rate could be threading dislocations \cite{chen1998pit},
inversion domains \cite{lucznik2009bulk}, cracks, substrate contamination \cite{sheftal1966magometov},
dielectric mask on a substrate\cite{motoki2007dislocation} or foreign particles landing on the film.
What kind of defects is responsible for pit formation in a particular case depends on the growth parameters, condition of the substrate and the reactor design.

 Pits could be  also developed as a result of incomplete coalescence of non-continuous films \cite{wagner2002influence}.

Sometimes pits can be healed via the overgrowth process.
Two types of the pit overgrowth process have been observed.
 \begin{figure}[]%
 \includegraphics*[width=1.0\linewidth]{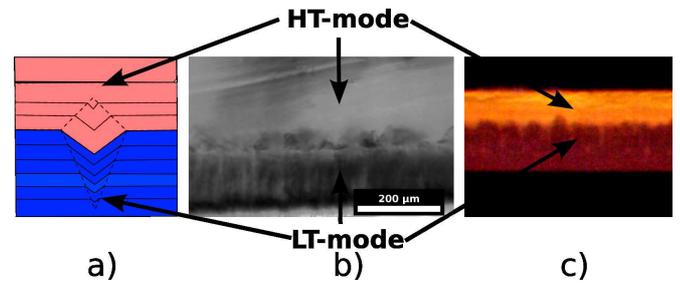}
 \caption[]{Pit overgrowth caused by transition from LT to HT mode.
 a) A schematic representation of the process.
The black lines represent the surface of the film at different moments of growth.
Dashed lines represent the trace left by the pit.
The pit increases in size during growth in LT mode.
In HT mode the growth rate of pit facets increases and the pit overgrows. 
b) and  c) are transmission light and photoluminescence micrographs of films grown in two stages.
Numerous pits formed during LT-mode growth were entirely overgrown when growth was switched to HT conditions.
 }
 \label{og1}
 \end{figure}

\begin{figure}[]%
 \includegraphics*[width=1.0\linewidth]{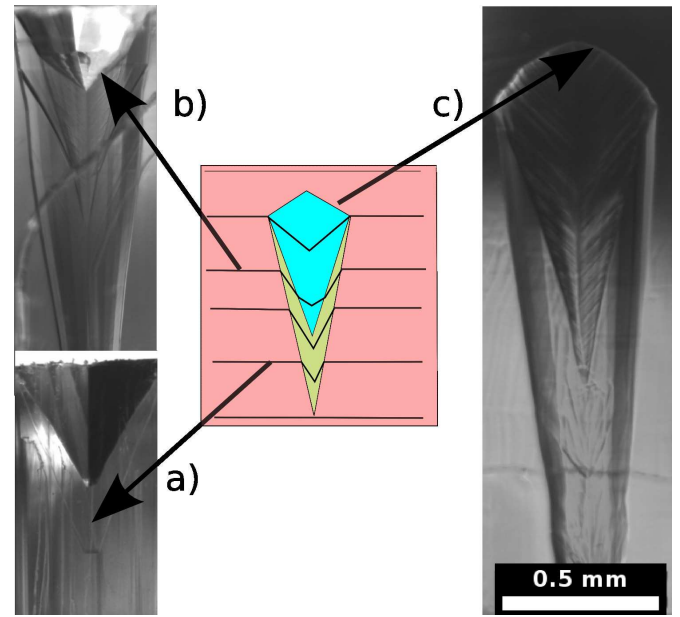}
 \caption[]{The stages of spontaneous pit overgrowth:
a) pit formed by slow facets increases in size.
b) fast growing facets develop on the bottom of the pit.
c) when the hole pit is covered by fast facets it is overgrown.
  }
 \label{og2}
 \end{figure}

This first type  of overgrowth was observed when the temperature was increased during growth process and the growth mode changed from LT to HT.
The growth rate of pit facets in HT mode is higher than the growth rate of \textit{c}-plane, and the pit is overgrown, Fig.~\ref{og1}a.
As a result, all pits formed in LT-mode overgrow entirely, Fig \ref{og1}b,c.

Another type of overgrowth occurs spontaneously.
The different stages of this process are depicted in Fig.~\ref{og2}.
A fast-growing facet is nucleated at the bottom of pit in the beginning of the process.
This facet is increased in size until it occupies the entire pit.
After that the  pit starts to shrink until complete overgrowth.

\begin{figure}[]%
\begin{center}
 \includegraphics*[width=0.96\linewidth]{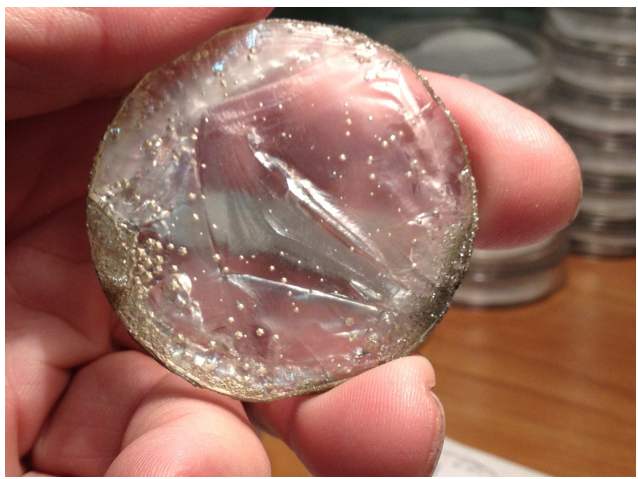}
\end{center}
 \caption[]{Photograph of a free standing  GaN boule with thickness of 3 mm obtained by the two-stage growth process.
  }
 \label{30cm}
 \end{figure}

Films grown in single stage at LT or HT mode had certain macrodefects: cracks for HT mode, pits for LT mode.
The two-stage method \cite{tavernier2002progress} was applied in order to  overcome the problem.
Crack-free films with thickness up to 3 mm and low density of the pits were grown,~Fig.~\ref{30cm}.
In this case the  first stage was conducted in LT-mode to provide low stress at the beginning of growth.
However, the  films grown during the first stage had rough surfaces with numerous pits.
To overgrow these pits and achieve smooth surface the second stage was conducted in HT mode.
The pits, formed at LT mode, were overgrown simultaneously when growth parameters have been switched to HT mode.
HT mode films grown on top of LT mode films were crack-free, unlike HT films grown directly on sapphire.

\paragraph*{Proposed model}
To explain the observed difference between the growth modes we suggest that the main  difference between LT and HT modes could be the rate of two-dimensional nucleation.
 The nucleation rate is low in HT mode.
The growth in this case proceeds by step flow, with steps provided by substrate misorientaition and screw dislocations.
Decreasing temperature or increasing  P$_{ \rm GaCl}$ increases the concentration of adsorbed species.
When the critical concentration is reached, the nucleation rate rapidly increases.
Therefore, the dashed line in Fig.~\ref{modes}a represents the onset of rapid surface nucleation.
Large density of steps on the rough surface slows down the surface diffusion of point defects \cite{schwoebel1966step},
and, thus, reduces dislocation climb~\cite{cantu2005role} and the related growth stress in LT growth mode.

\section{Conclusions}

Two growth modes (LT and HT) were observed in HVPE growth of GaN.
Films grown in these modes differed by surface roughness, density and shape of the pits and the value of growth stress.
A transition between modes took place in a very narrow area of process parameters.
Surface roughness, density and shape of the pits and the value of growth stress change simultaneously when the  growth mode is changed  from LT to HT.
The possible mechanisms for  transition between growth modes and generation of growth stress are proposed.

%
%
\providecommand{\WileyBibTextsc}{}
\let\textsc\WileyBibTextsc
\providecommand{\othercit}{}
\providecommand{\jr}[1]{#1}
\providecommand{\etal}{~et~al.}

\bibliographystyle{IEEEtran}
\bibliography{two-modes}

\end{document}